\documentclass[a4paper,fleqn]{cas-sc}

\usepackage[numbers]{natbib}

\def\tsc#1{\csdef{#1}{\textsc{\lowercase{#1}}\xspace}}
\tsc{WGM}
\tsc{QE}

\usepackage{siunitx}
\usepackage{tabularx}
\usepackage{here}

\begin{document}
\let\WriteBookmarks\relax
\def\floatpagepagefraction{1}
\def\textpagefraction{.001}

\shorttitle{Development of an Extensible Unified Control System}    

\shortauthors{R. Nishimura et al.}  

\title [mode = title]{Development of an Extensible Unified Control System Using the STARS Framework and Common Commands for Detector Control}  



\author[1,2]{Ryutaro NISHIMURA}[orcid=0000-0002-3156-7347]
\cormark[1]
\ead{ryutaro.nishimura@kek.jp}
\credit{Conceptualization of this study, Data curation, Formal analysis, Investigation, Methodology, Software, Visualization, Writing - original draft preparation, Writing - review and editing}

\author[1,2]{Yuki SHIBAZAKI}[]
\credit{Data curation, Funding acquisition, Investigation, Resources, Supervision, Validation, Writing - review and editing}

\author[1,2]{Daisuke WAKABAYASHI}[]
\credit{Data curation, Funding acquisition, Investigation, Resources, Supervision, Validation, Writing - review and editing}

\author[1]{Yoshio SUZUKI}[]
\credit{Data curation, Supervision, Validation, Visualization, Writing - review and editing}

\author[1,2,3]{Keiichi HIRANO}[]
\credit{Data curation, Funding acquisition, Resources, Supervision, Validation, Writing - review and editing}

\author[1]{Hiroaki NITANI}[]
\credit{Data curation, Funding acquisition, Investigation, Project administration, Resources, Supervision, Validation, Writing - review and editing}

\author[1]{Takashi KOSUGE}[]
\credit{Data curation, Funding acquisition, Investigation, Resources, Supervision, Validation, Writing - review and editing}

\author[1,2]{Noriyuki IGARASHI}[]
\credit{Data curation, Funding acquisition, Investigation, Project administration, Resources, Supervision, Validation, Writing - review and editing}

\affiliation[1]{organization={Institute of Materials Structure Science, High Energy Accelerator Research Organization (KEK)},
            addressline={}, 
            city={Tsukuba},
            postcode={305-0801}, 
            state={Ibaraki},
            country={Japan}}
\affiliation[2]{organization={Materials Structure Science Program, Graduate Institute for Advanced Studies, Graduate University for Advanced Studies (SOKENDAI)},
            addressline={}, 
            city={Tsukuba},
            postcode={305-0801}, 
            state={Ibaraki},
            country={Japan}}
\affiliation[3]{organization={Graduate School of Pure and Applied Sciences},
            addressline={}, 
            city={Tsukuba},
            postcode={305-8571}, 
            state={Ibaraki},
            country={Japan}}

\cortext[1]{Corresponding author}

\fntext[1]{}


\begin{abstract}
A zooming optical system comprising two Fresnel zone plates (FZPs) was developed and installed at the AR-NE1A beamline of the Photon Factory, High Energy Accelerator Research Organization (KEK), Japan. 
To ensure reliable and versatile operation, we implemented a dedicated control architecture based on the Simple Transmission and Retrieval System (STARS) framework and the newly proposed STARS Common Commands for Detector Control (CCDC)---a data-acquisition (DAQ) state model and command set designed specifically for detector control. 
The system serves both as a practical control system for the zooming optics and as a demonstration of modular extensibility using STARS and detector interoperability through CCDC. 
The system has been commissioned, and its performance has been verified at the AR-NE1A beamline. 
The architecture enables flexible configuration of optical components and provides a unified interface for both routine operation and advanced experimental protocols. 
\end{abstract}


\begin{highlights}
\item Unified control system developed for 2-FZPs zooming optics. 
\item Proposed system based on STARS framework and common commands for detector control.
\item The control system was tested at the AR-NE1A beamline at KEK, Japan. 
\end{highlights}

\begin{keywords}
Beamline Control \sep Zooming Optics \sep Fresnel zone plate \sep X-ray microscopy \sep STARS \sep DAQ
\end{keywords}

\maketitle

\section{Introduction}

Synchrotron radiation facilities, including the Photon Factory (PF) at the High Energy Accelerator Research Organization (KEK) in Japan, continually advance their optical systems to meet the growing demand for high-precision X-ray experiments. 
As these optical systems become increasingly complex, the focus has gradually shifted from individual control systems that require specialist expertise toward reliable, user-friendly integrated control systems that ensure stable operation under a wider range of experimental conditions. 
However, the human resources available to maintain these systems are limited. 
Control systems must therefore accommodate diverse equipment and experimental methods while reducing maintenance effort, promoting standardization where feasible, and adopting modular designs. 
These features facilitate system reconfiguration and reuse. 
Control frameworks that address these requirements include
DAQ-Middleware \cite{daq-middleware1, daq-middleware2}, which is mainly used in neutron and muon experiments at J-PARC's Materials and Life Science Experimental Facility (MLF) \cite{j-parc1, j-parc-mlf1}; 
MADOCA \cite{madoca1}, which is mainly used at SPring-8 \cite{spring8-1} and NanoTerasu \cite{nanoterasu-1}; 
EPICS \cite{epics1, epics2}, which is primarily used to control large accelerators at facilities such as J-PARC, SuperKEKB \cite{superkekb1}, and LANSCE \cite{lansce1}; 
TANGO \cite{tango1}, which is primarily used at European synchrotron radiation facilities such as ESRF \cite{esrf1}; and 
TINE \cite{tine1}, which is primarily used for accelerator control at DESY \cite{desy1}. 
At the PF beamlines, the Simple Transmission and Retrieval System (STARS) framework \cite{Stars1, Stars2} has been used primarily. 
Although their implementations vary according to facility-specific requirements, these frameworks share a common structure in which modules are divided by function or device and interconnected through communication protocols such as TCP/IP. 
This structure facilitates reconfiguration and the reuse of developed modules, thereby providing high scalability. 
It also enables development assets to be shared within and across facilities. 
The methods used to link modules differ among frameworks and can be broadly classified as stateful or stateless: 
stateful designs manage all modules using common commands and shared data acquisition (DAQ) states, 
whereas stateless designs allow each module to use its own commands without sharing common DAQ states. 
In stateful designs, module-control mechanisms are standardized, allowing differences among devices to be accommodated within the framework. 
However, strict DAQ-state management and precise control of state transition sequences among modules are required; 
therefore, stateful designs tend to be less scalable than stateless designs. 
STARS uses a stateless design because it places minimal requirements on the sharing of a common DAQ state among beamline components. 
This design prioritizes flexibility and facilitates frequent modifications to experimental apparatus and measurement methods. 
Its relatively low design cost also makes it advantageous for the continued development of beamline systems. 
However, device-specific control requirements are embedded in the module commands. Consequently, replacing a device with a similar one may require extensive modification of the control system.
To address this limitation, we propose introducing dedicated DAQ states for experimental devices that are expected to be replaced frequently. 
This approach enables basic control through common STARS commands and facilitates the replacement and use of devices of the same type that have different characteristics. 
It therefore allows measurement systems to better utilize the scalability of STARS, which is already widely deployed at the PF. 

Although framework-specific device abstraction mechanisms commonly provide interoperability, maintaining these mechanisms may require development and maintenance at the framework level. 
Therefore, a lightweight approach that improves device interoperability while preserving the simplicity and flexibility of the underlying framework is desirable. 

Unlike control frameworks that achieve interoperability through middleware-level device abstractions, the present approach preserves the lightweight message-relay architecture of STARS and realizes detector interoperability through the standardization of detector-control commands and DAQ states at the client level. 
This architecture enables detector interchangeability without requiring modification of the STARS server or the underlying framework itself. 

We developed a unified control framework as a model implementation of this approach using dedicated DAQ states. 
This system demonstrates modular extensibility within the STARS framework and detector interoperability through the proposed STARS Common Commands for Detector Control (CCDC), a common DAQ-state model and command set designed for detector control. 
It also provides flexibility for changes in optical configuration and an intuitive interface for both routine users and advanced experimenters. 
This system was developed as a practical control system for the zooming optics, which comprise two Fresnel zone plates (FZPs) \cite{Wakabayashi1} and are installed at the PF AR-NE1A beamline. In this study, we evaluated the proposed system through automated measurement tests that combined STARS-based control of the 2-FZPs zooming optics with CCDC-based detector control. 

\section{Overview of the Zooming X-ray Microscope at AR-NE1A}

\subsection{2-FZPs Zooming Optics}

\begin{figure}
\centering
\includegraphics[width=\linewidth]{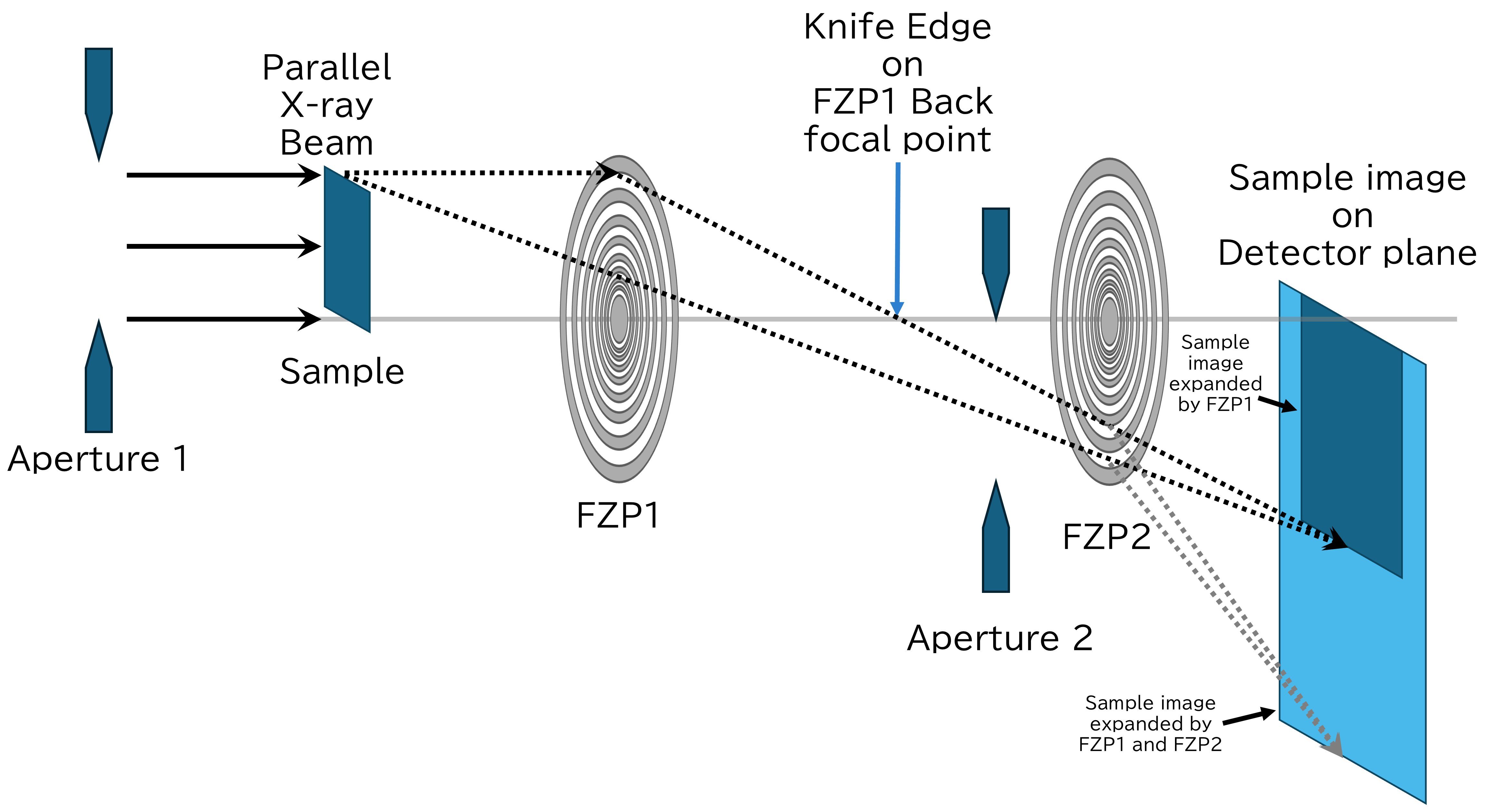}
\caption{\label{fig_zoomingoptics_schem} Schematic of the zooming optical system comprising two Fresnel zone plates (FZPs). In this system, adjusting the position of FZP2, including its removal from the optical path, enables a wide magnification range even with a short camera length. The low-magnification mode using only FZP1 is termed the 1-FZP mode, whereas the high-magnification mode using both FZP1 and FZP2 is termed the 2-FZP mode. Furthermore, inserting a knife edge at the back focal plane of FZP1 enables the acquisition of a schlieren phase-contrast image.}
\end{figure}

The 2-FZPs zooming optical system \cite{Wakabayashi1} (Figure \ref{fig_zoomingoptics_schem}) enables variable-magnification X-ray microscopy by adjusting the FZP positions while maintaining a fixed camera length, defined here as the object-to-camera distance. 
This system provides a wide magnification range of approximately \(25\times\)--\(300\times\) with a relatively short camera length of approximately 6.9 m and was initially installed at the PF AR-NE1A beamline \cite{Ar-ne1a}. 
The X-ray beam from a multipole wiggler is monochromatized using a Si(111) double-crystal monochromator and collimated by variable-bent double multilayer-mirror optics in a crossed-mirror geometry (Kirkpatrick--Baez optics) to produce parallel illumination. 
After passing through the aperture, the beam illuminates the sample, with only the lower half of FZP1 irradiated for diffraction-order sorting. 
After passing through the sample, the X-ray beam is magnified by FZP1 and, in the high-magnification mode, is further magnified by FZP2. 
The FZPs, manufactured by Applied Nanotools, Inc., have a diameter of 550 \si{\micro\meter} and Au zones more than 600 nm thick. 
The outermost zone widths \(\Delta {r}_{n}\) of FZP1 and FZP2 are 25 and 50 nm, respectively. 
For schlieren phase-contrast imaging, a knife edge positioned at the back focal plane of FZP1 partially blocks the beam. 
The beam then passes through a 5-m vacuum pipe before reaching the X-ray camera. 

\subsection{Operational Requirements for Optical Control}
Although the 2-FZPs zooming optical system has a relatively simple configuration, changing the magnification requires adjustment of the FZPs, apertures, knife edges, and other optical elements. 
The sample-stage position must also be adjusted to meet the measurement requirements. 
Controlling the positions of these elements requires the operation of up to 32 motorized stages. 
Depending on the desired magnification and sample type, changing the X-ray energy also requires adjustment of the upstream monochromator and multilayer mirrors, which are controlled by up to 16 motorized stages. 
Therefore, a control system with a user-friendly interface is needed to enable operation by users without specialized beamline expertise. 

\subsection{Detector Selection}
The X-ray energy used in this optical system depends on the desired magnification and sample type. 
The required field of view, resolution, and other measurement parameters also vary among target samples. 
Because a single detector cannot accommodate all sample types and measurement conditions, different detectors must be used. 
Therefore, the control system must support detector control and efficient switching between detectors.

\section{Proposed New Control System}

To address these requirements, we developed a new control system. 
This section describes the key technologies used in the system and provides an overview of its architecture. 

\subsection{STARS Framework}

STARS \cite{Stars1, Stars2} is a message-transfer framework for equipment control at the PF beamlines. 
Designed for small-scale control systems based on standard TCP/IP sockets, the framework can be used with various operating systems. 
It also supports the development of control and measurement modules using various programming languages capable of handling TCP/IP sockets. 

\begin{figure}
\centering
\includegraphics[width=\linewidth]{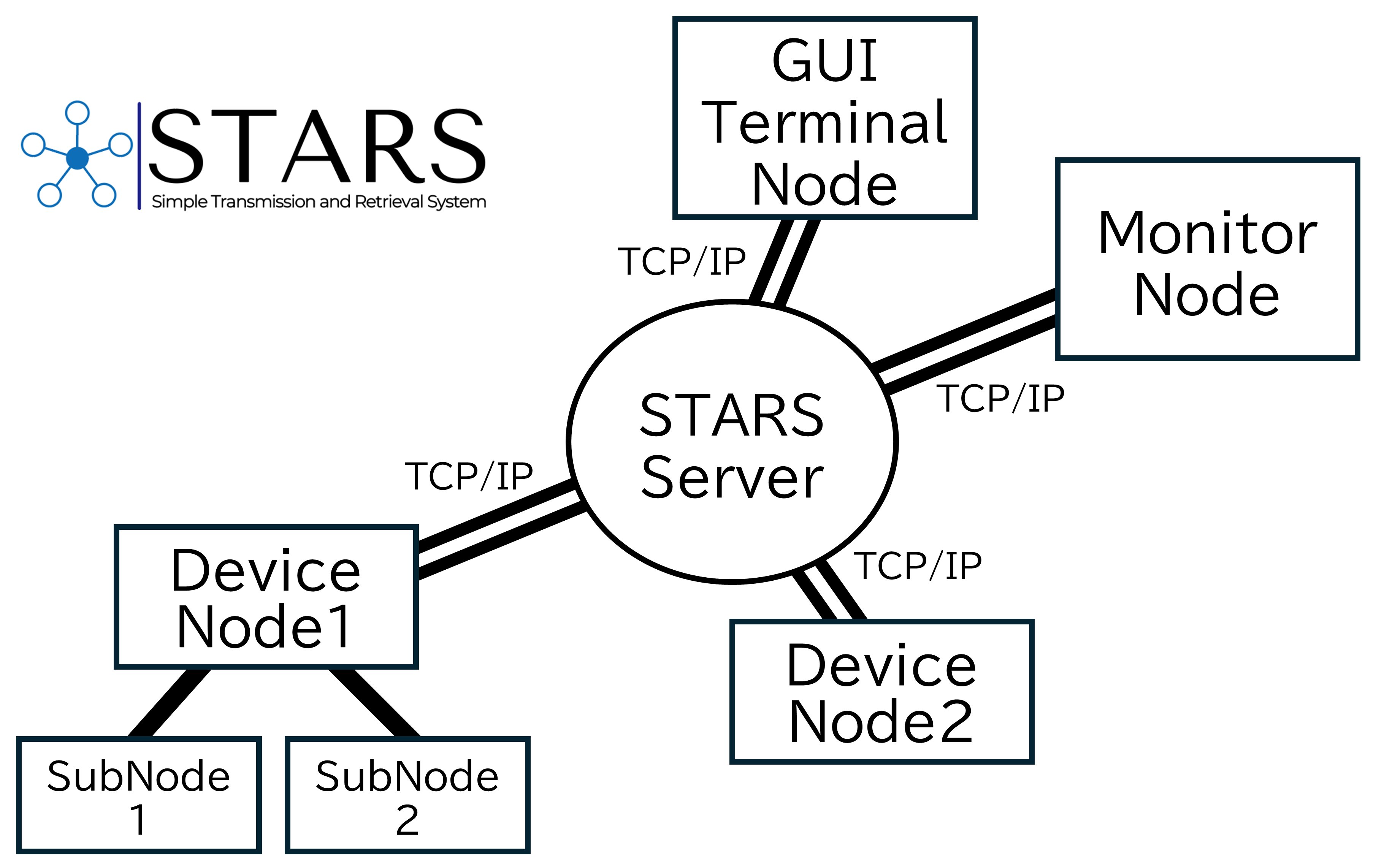}
\caption{\label{fig_stars_schem} Schematic of the STARS framework. The framework consists of client programs called STARS clients (shown as Device Node 1, Device Node 2, GUI Terminal Node, and Monitor Node) and a server program called the STARS server, shown at the center of the figure. Each STARS client connects to the STARS server via TCP/IP.}
\end{figure}

The framework consists of client programs, referred to as STARS clients, and a server program, referred to as the STARS server. Each STARS client connects to the STARS server via TCP/IP (Figure \ref{fig_stars_schem}). 
The STARS clients and server exchange only text-based messages, which 
can support structured data formatted in JavaScript Object Notation (JSON), Extensible Markup Language (XML), and plain text.
Each STARS client has a unique node name and sends messages to the server by specifying the destination node name.
STARS clients may also have subnodes, and commands can be sent directly between nodes by specifying the destination. This functionality facilitates the construction of hierarchical systems.

\subsection{STARS CCDC}

\begin{figure}
\centering
\includegraphics[width=\linewidth]{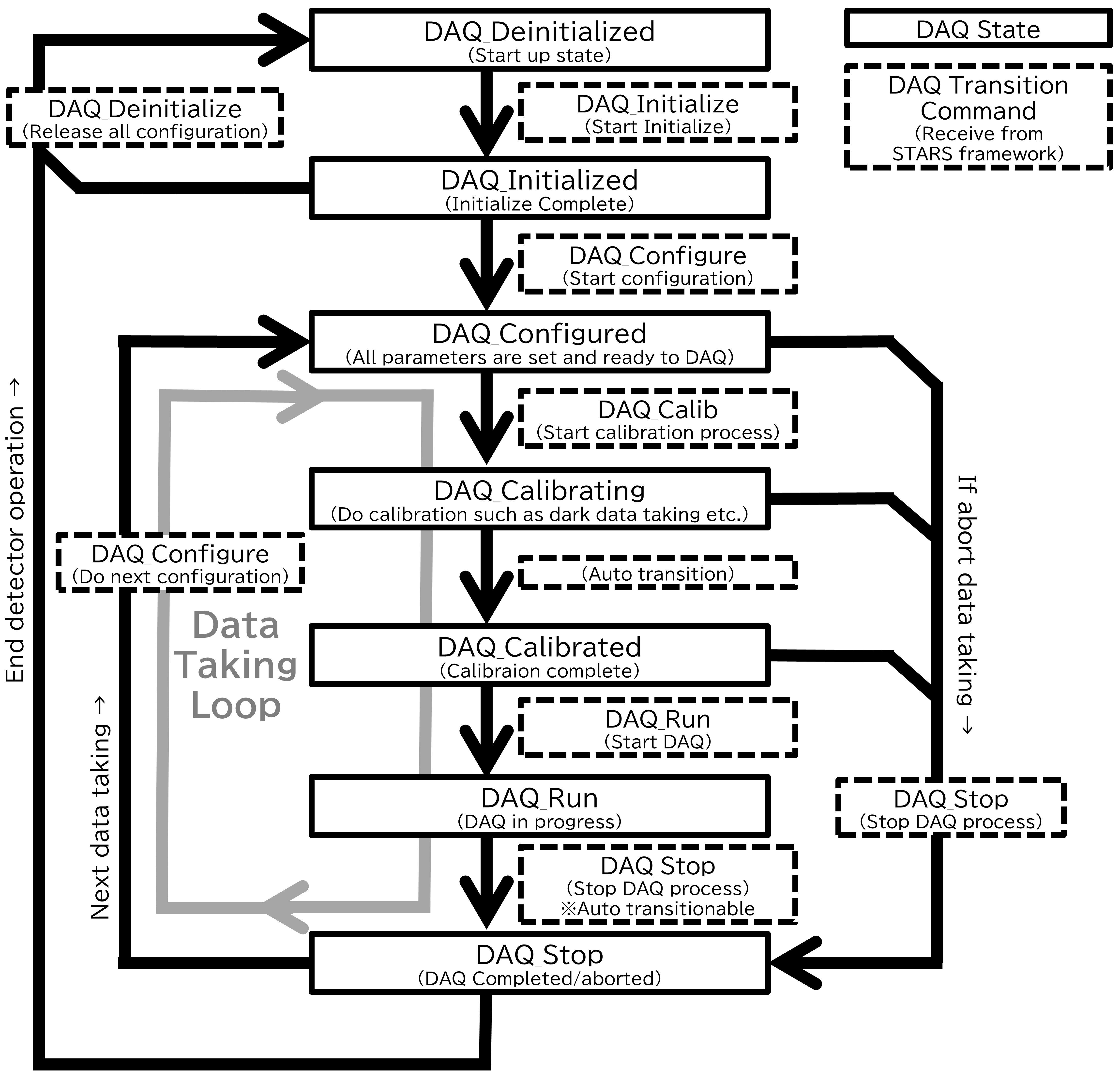}
\caption{\label{fig_cdcc_schem} Schematic of STARS CCDC within a detector-control STARS client. A CCDC-compatible detector-control client implements six DAQ transition commands and seven DAQ states. The states represent the following conditions: DAQ\_Deinitialized -- initial state upon software launch; DAQ\_Initialized -- initialization complete; DAQ\_Configured -- parameters required for detector operation configured; DAQ\_Calibrating -- detector calibration in progress; DAQ\_Calibrated -- detector calibration complete and operational measurements ready to proceed; DAQ\_Run -- detector measurement in progress; and DAQ\_Stop -- measurement stopped.}
\end{figure}

STARS CCDC \cite{StarsCcdc} is a detector-control command system designed to provide basic detector control within STARS. 
It improves interoperability between detectors, enabling detectors suited to specific measurement conditions to be selected without major modifications to the overall measurement system. 

Unlike control frameworks that provide interoperability through middleware-level device abstractions, STARS maintains a lightweight message-relay architecture. 
CCDC improves detector interoperability while preserving this architectural simplicity by introducing common detector-control commands and DAQ states at the client level without requiring modification to the STARS server implementation. 

In X-ray imaging, detector selection may depend on the sample, optical system, and X-ray energy. 
Because the control commands and parameters are specific to each detector, they are fundamentally incompatible across detectors. 
Consequently, adapting automated measurement software that also controls the optical system and peripheral equipment for each detector is challenging. 
CCDC addresses this issue by abstracting the minimum set of control commands and transmission parameters required to acquire a single dataset (e.g., one image), rather than directly exposing detector-specific commands and parameters within the detector-control STARS client. 
These detector-specific commands and parameters are encapsulated within the client, enabling detectors to be switched without modifying the measurement-control system. 
This abstraction is implemented using the seven DAQ states shown in Figure \ref{fig_cdcc_schem} and six DAQ transition commands that control transitions among these states. 
When a CCDC-compatible detector-control client is launched, it initially enters the ``DAQ\_Deinitialized'' state.
\begin{description}
\item[DAQ\_Deinitialized]
Initial state upon software launch. 
In this state, no detector-specific preparation, including establishment of a hardware connection, has been completed. 
\item[DAQ\_Initialized]
State reached after initialization. 
In this state, a connection to the detector hardware has been established. 
\item[DAQ\_Configured]
State in which the parameter settings required for detector operation have been completed. 
\item[DAQ\_Calibrating]
State in which the detector calibration is in progress. 
\item[DAQ\_Calibrated]
State indicating that the detector calibration is complete and measurements can proceed. 
\item[DAQ\_Run]
State in which measurement by the detector is in progress. 
\item[DAQ\_Stop]
State indicating that the measurement has stopped. 
\end{description}

In CCDC, DAQ-state transitions are normally triggered by DAQ transition commands received from STARS. 
The transitions from DAQ\_Calibrating to DAQ\_Calibrated and from DAQ\_Run to DAQ\_Stop may occur automatically when the corresponding detector-side processes are completed. 
If an invalid DAQ transition is requested or a transition fails for any reason, the detector-control client returns an error response according to the standard STARS protocol and remains in its current DAQ state. 
Timeout handling is intentionally delegated to the detector-control client implementation because initialization, calibration, and acquisition times vary substantially among detector systems. 
This design preserves the simplicity and extensibility of the CCDC specification while maintaining compatibility with diverse detector implementations. 

In normal operation, the transition from DAQ\_Run to DAQ\_Stop is performed automatically when data acquisition has been completed. 
However, a DAQ\_Stop transition command may also be issued externally to abort an ongoing acquisition or to recover from abnormal conditions. 
Because the DAQ states defined by CCDC are handled entirely within the detector-control client and do not directly affect the operation of the STARS framework, failures occurring within the detector-control client are generally isolated from the rest of the control system. 
Therefore, recovery may also be achieved by restarting the detector-control client. 

A single data-acquisition cycle proceeds through the DAQ states from DAQ\_Configured to DAQ\_Stop (the ``Data-taking loop'' in Figure \ref{fig_cdcc_schem}). 
To acquire multiple datasets, the system repeats this state-transition cycle the required number of times. 
The DAQ states defined by CCDC are intended to enable mutual interpretation of the DAQ states and commands when the detector-control software must operate across multiple frameworks. 

When developing CCDC-compatible STARS clients for individual detectors, 
developers must organize detector-specific commands and parameters into categories that should be executed and transmitted at each DAQ state. 
Table \ref{tab_ccdc_soipix_mapping} provides an example mapping between the CCDC DAQ states and detector-specific actions implemented for the INTPIX4NA SOIPIX detector. 
During DAQ-state transitions, the detector-control client executes the detector-specific commands associated with each DAQ state while exposing only the standardized CCDC interface to the upper-level measurement-control software. 
In this way, detector-specific control procedures are encapsulated within the detector-control client, enabling detector interchangeability without modifying the overarching measurement-control system. 
This detector-control client must implement this categorization so that commands are processed appropriately according to the DAQ-state transitions. 
To maintain interoperability, the DAQ states and transition commands are fixed and cannot be changed. 
However, other commands, such as those used to set detector-specific parameters, may be implemented as needed. 
Developers must ensure that measurements based on DAQ transition commands remain executable after the detector parameters have been configured in the client. 
For certain parameters, such as exposure time, a recommended command name is provided. 

DAQ software developed at KEK currently supports CCDC for the Hamamatsu C12849-111U sCMOS detector \cite{hpkscmosc}, the INTPIX4NA SOIPIX detector \cite{intpix4na1, intpix4na2}, the GPixel GSENSE400BSI-PS-based soft X-ray sCMOS detector \cite{artraygsense400bsi}, and the Dectris Eiger2X detector \cite{dectris_eiger2}. 
The same software architecture can be extended to additional detectors in the future. 
By consolidating detector operations into a common set of DAQ transition commands, CCDC reduces the extent of detector-specific modification required in measurement procedures. 
This architecture also simplifies the integration of additional detector systems, provided that a CCDC-compatible detector-control client is implemented. 

\begin{table}[H]
 \caption{Example mapping between CCDC DAQ states and detector-specific actions or statuses for the INTPIX4NA SOIPIX detector.}
 \begin{tabularx}{\linewidth}{p{8.5em}X}
 \hline
 CCDC DAQ State & Example detector-specific action or status (SOIPIX) \\
 \hline
 DAQ\_Deinitialized & No detector connection established \\
 DAQ\_Initialized & Open the detector connection and initialize the hardware \\
 DAQ\_Configured & Apply all detector parameters (e.g., exposure time, total number of frames, detector-circuit operating parameters, trigger mode and other acquisition settings) and prepare data storage on the DAQ PC for image-data collection \\
 DAQ\_Calibrating & Acquire background images for calibration \\
 DAQ\_Calibrated & Background-image acquisition is complete; the system is ready for image data acquisition \\
 DAQ\_Run & Start data acquisition, store the acquired data, and perform on-the-fly data processing, such as image integration \\
 DAQ\_Stop & Stop data acquisition and finalize the measurement \\
 \hline
 \end{tabularx}
 \label{tab_ccdc_soipix_mapping}
\end{table}

\subsection{Overview of the Proposed Control System}

\begin{figure}
\centering
\includegraphics[width=\linewidth]{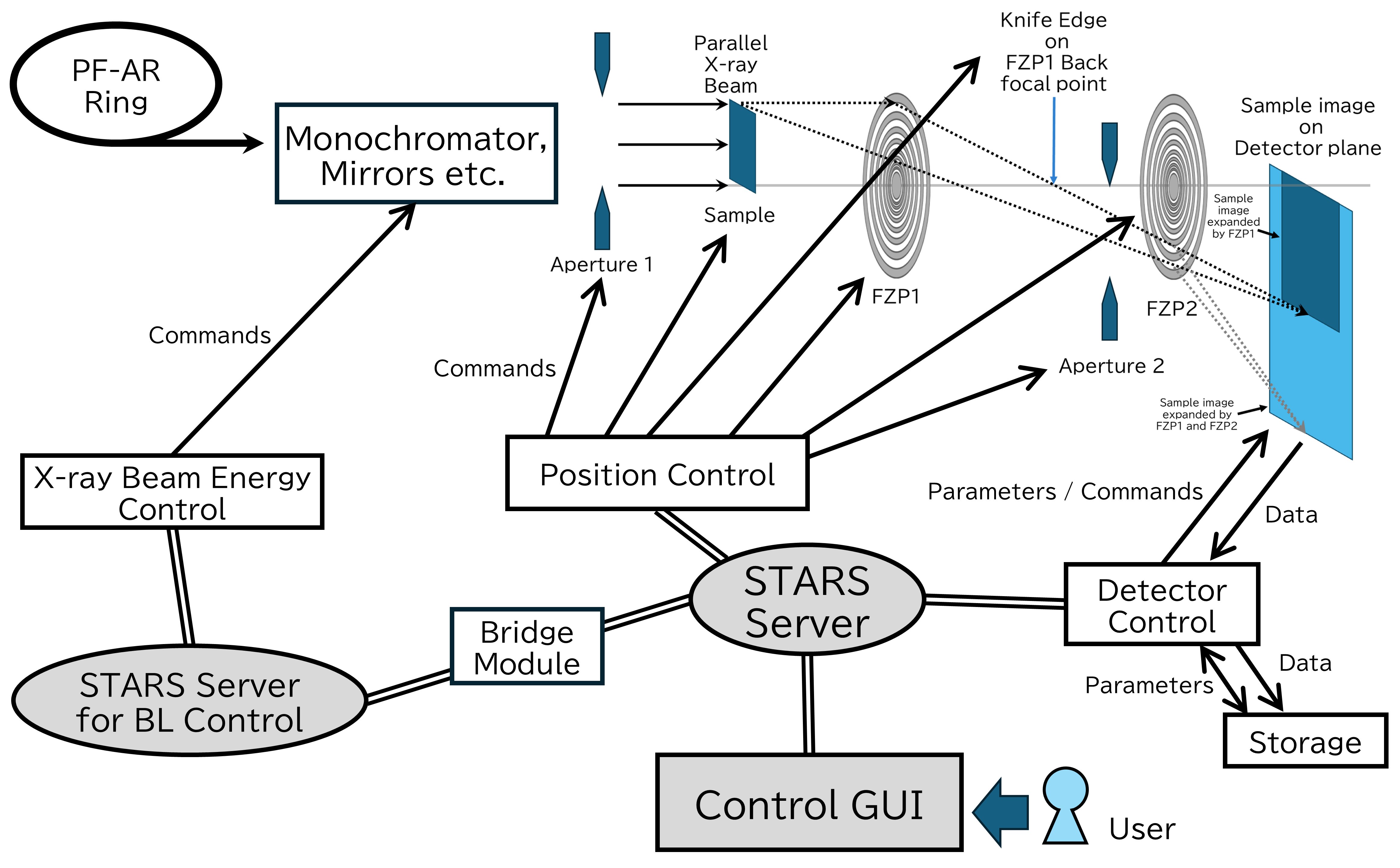}
\caption{\label{fig_zoomingctrl_schem} Schematic of the proposed control system. The optical system and detector are controlled through independent modules that function as STARS clients and are interconnected through the STARS server. 
The system is also connected through a relay to the existing STARS-based control system for upstream components, such as monochromators and mirrors, required for X-ray energy control. Users control the system and perform measurements through a unified control GUI.}
\end{figure}

The proposed control system was constructed using the STARS framework and CCDC. 
Figure \ref{fig_zoomingctrl_schem} presents a schematic of the system architecture. 
In this system, control of the optical system, involving up to 32 motorized stages, and detector control are implemented as independent modules that function as STARS clients. 
These modules are interconnected through the STARS server. 
The system is also connected through a relay to the existing STARS-based control system for upstream components, such as monochromators and mirrors, that are required for X-ray energy control and are operated using up to 16 motorized stages. 
Users can control the setup and perform measurements through a unified control graphical user interface (GUI) module, which enables integrated control of the existing control systems. 
This unified control GUI module incorporates various adjustments and automated measurement algorithms described below and is implemented using an object-oriented structure intended to facilitate its partial or complete adaptation to other systems in the future.

In this system, users can select CCDC-compatible detectors according to the experimental conditions. 
The Hamamatsu sCMOS detector C12849-111U \cite{hpkscmosc} and the INTPIX4NA SOIPIX detector \cite{intpix4na1, intpix4na2} are currently available for use. 
The Hamamatsu sCMOS detector C12849-111U has a pixel format of \(2048 \times 2048\), a pixel size of 6.5 {\textmu}m, a 10-{\textmu}m-thick gadolinium oxysulfide (P43) scintillator, and 1:1 fiber optics. 
Because it employs a scintillator, it is an indirect-conversion detector and offers lower resolution than direct-conversion detectors. 
However, the scintillator can detect high-energy X-rays, which have high penetrating power, and the detector electronics are relatively resistant to radiation damage. Consequently, it is well suited to imaging under conditions of relatively high X-ray intensity and energy, such as 14.4 keV in this optical system. 
The INTPIX4NA SOIPIX detector has a pixel format of \(832 \times 512\), a pixel size of 17 {\textmu}m, and a 320-{\textmu}m-thick high-resistivity silicon substrate for X-ray detection. 
As a direct-conversion detector, it provides efficient X-ray detection and high sensitivity and excellent resolution. 
It is particularly well suited to imaging under conditions of relatively low X-ray intensity and energy, such as 9.6 keV in this optical system. 
This system moves the optical and upstream components to predetermined positions using precalibrated preset values according to the X-ray energy selected by the user through the GUI. 
This adjustment enables users to begin measurements promptly without spending considerable time or effort readjusting the energy and optics for each experiment. 
The system also provides several semiautomated control and measurement functions, as described below.

\begin{description}

\item[Energy and Optics Switching]
This function enables semiautomatic switching of the X-ray energy and tuning of the zooming optics using preset values. 
This function is required for the following reasons. 
\begin{enumerate}
\item The magnification provided by the optical system depends on the X-ray energy. 
At 14.4 keV, the magnification ranges from approximately \(25\times\) to \(150\times\), 
whereas at 9.6 keV, it exceeds approximately \(300\times\) \cite{Wakabayashi1}. 
High X-ray energies provide lower magnification, whereas low energies provide higher magnification. 
Switching the X-ray energy therefore enables observations 
over a magnification range that cannot be achieved using a single X-ray energy. 
\item The required X-ray energy also depends on the sample and measurement conditions. Samples with high absorption, such as thick samples or those containing heavy elements, 
require high-energy X-rays with greater penetrating power. 
Conversely, thin samples observed at maximum magnification require low-energy X-rays, 
even when the transmission efficiency is reduced. 
\end{enumerate}

\item[One-Shot Imaging] 
This function provides a simple method for acquiring a single image without peripheral control. 
It sequentially issues commands to cycle through the CCDC DAQ states. 
The resulting detector behavior depends on the implementation of the detector-side software. 

\item[Focus Check] 
This measurement protocol acquires data for adjusting the focus of the optical system. 
The function moves the FZP or sample in predetermined increments, acquires an image at each position, and records the corresponding positional information and images. 

\item[Two-Dimensional Multishot] 
In this measurement, the sample is moved repeatedly in predetermined increments, and an image is acquired at each position. 
This function enables the imaging of samples larger than the standard field of view. 
Users can independently specify the displacement and number of steps for the horizontal and vertical movements of the sample. 
The acquired images can then be stitched into a single large image using post-processing software. 

\item[Computed Tomography / Laminography Data Acquisition] 
This measurement protocol acquires datasets for computed tomography or laminography.
An X-ray profile image, denoted $I_0$, is first acquired for normalization. 
After the sample has been positioned, it is rotated in predetermined angular increments, and an image is acquired at each position, producing a complete dataset spanning \ang{180} or \ang{360}. 

\end{description}

\section{Demonstration Measurements}

A set of measurements was performed using the developed control system. 
These measurements were conducted primarily to functionally validate the proposed control architecture rather than to characterize the imaging performance of the optical system. 
The X-ray microscope was installed at the PF AR-NE1A beamline at KEK. 
These measurements were performed using the Hamamatsu C12849-111U sCMOS detector and the INTPIX4NA SOIPIX detector. 
The demonstration samples were Siemens star and line-and-space test patterns from a 600-nm-thick Au calibration standard (CAL21HEI2, Applied Nanotools). 

\subsection{Energy and Optics Switching}

\begin{figure}
\centering
\includegraphics[width=\linewidth]{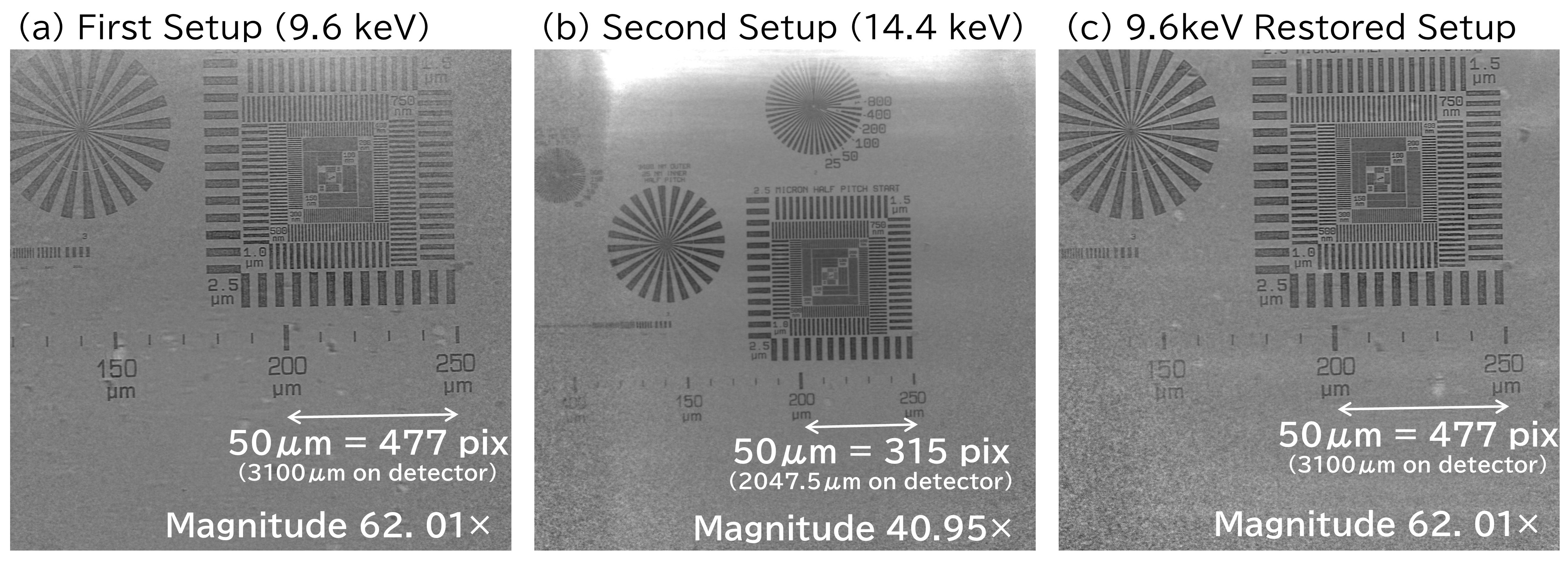}
\caption{\label{fig_mag_test} Results of the energy and optics switching test. Test-pattern images were acquired using the Hamamatsu C12849-111U sCMOS detector and normalized using the X-ray beam profile. (a) Initial monochromatic X-ray condition at 9.6 keV, with a magnification of \(61.01\times\). This condition was stored as a preset in the control system. (b) Monochromatic X-ray condition at 14.4 keV, with a magnification of \(40.95\times\). (c) Monochromatic X-ray condition at 9.6 keV restored by the control system after operation at 14.4 keV.}
\end{figure}

In this measurement, the X-ray energy and optical configuration were switched in the following sequence. 
First, the optical system was tuned for operation at an X-ray energy of 9.6 keV, and the resulting configuration was stored as a preset. 
Second, the energy was changed to 14.4 keV, and the optical system was retuned. 
Finally, the energy was returned to 9.6 keV using the initial preset to confirm restoration of the optical configuration. 
This measurement was performed in the 1-FZP mode. 
Figure \ref{fig_mag_test} shows the measurement results obtained using the Hamamatsu C12849-111U sCMOS detector. 
Figures \ref{fig_mag_test}(a), (b), and (c) show images of the test patterns acquired at 9.6 keV under the preset condition, at 14.4 keV, and at 9.6 keV after restoration of the preset configuration, respectively. 
Comparison of Figures \ref{fig_mag_test}(a) and (c) shows that the magnification and focus were reproduced after the optical configuration was restored. 
These results confirm the functionality of the optical-control component of the developed control system. 

\subsection{Two-Dimensional Multishot}

\begin{figure}
\centering
\includegraphics[width=\linewidth]{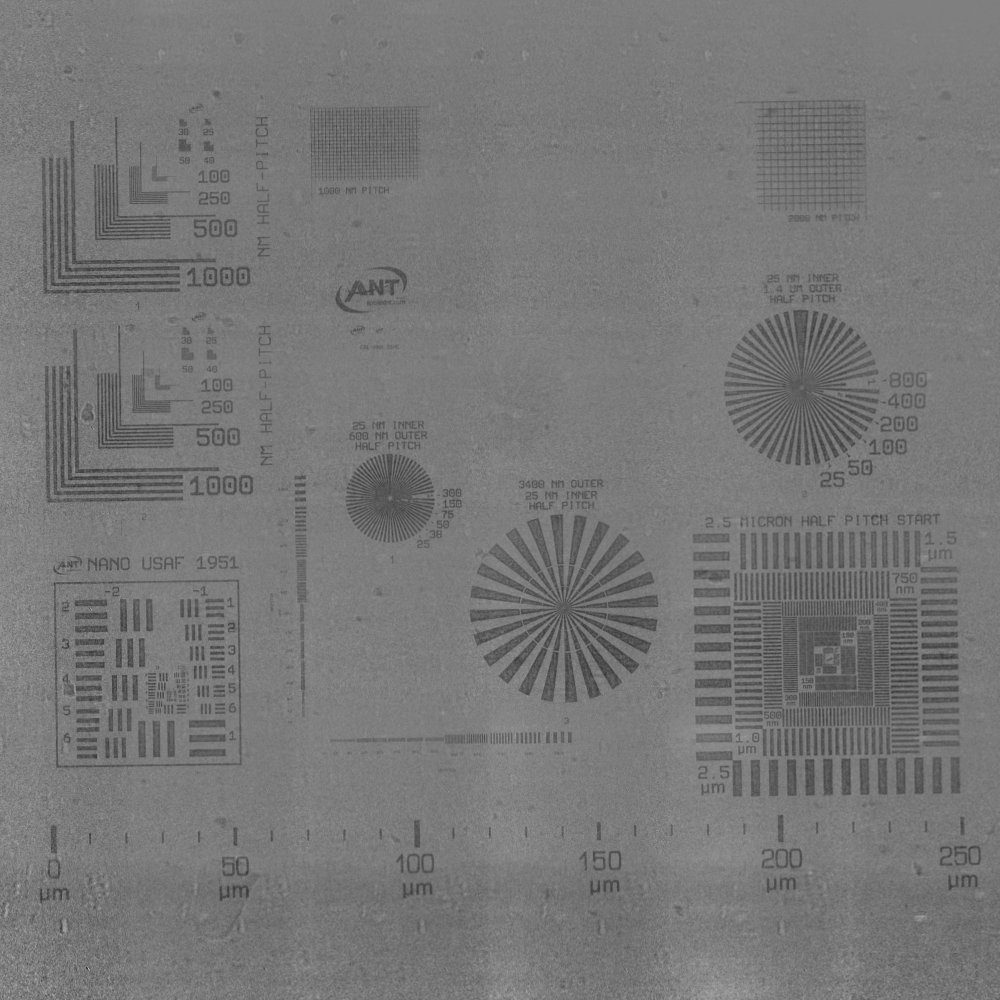}
\caption{\label{fig_two_dim_ms_soipix} Stitched image reconstructed from a \(5 \times 5\) two-dimensional multishot dataset acquired using the INTPIX4NA SOIPIX detector at an X-ray energy of 9.6 keV. The image was normalized using the X-ray beam profile and processed using the Grid/Collection stitching plugin in Fiji.}
\end{figure}

\begin{figure}
\centering
\includegraphics[width=\linewidth]{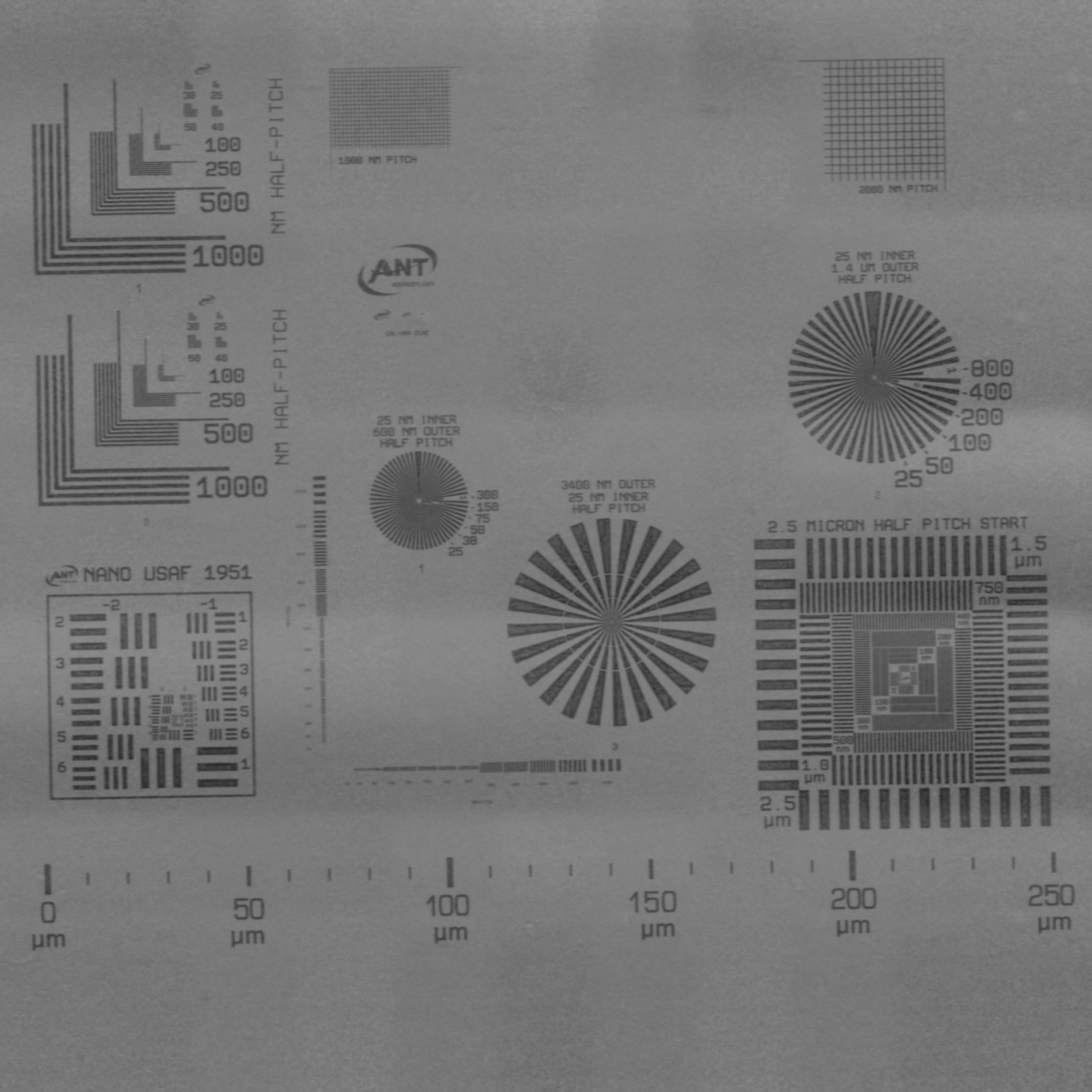}
\caption{\label{fig_two_dim_ms_scmos} Stitched image reconstructed from a \(5 \times 5\) two-dimensional multishot dataset acquired using the Hamamatsu C12849-111U sCMOS detector at an X-ray energy of 14.4 keV. The image was normalized using the X-ray beam profile and processed using the Grid/Collection stitching plugin in Fiji.}
\end{figure}

A continuous automated measurement was performed in the 1-FZP mode at 25 positions arranged in a \(5 \times 5\) grid with 50-{\textmu}m steps. The test pattern exceeded the field of view of the optical system. 
Figure \ref{fig_two_dim_ms_soipix} shows a stitched image of the test pattern reconstructed from the \(5 \times 5\) two-dimensional multishot dataset acquired using the INTPIX4NA SOIPIX detector at an X-ray energy of 9.6 keV. 
Figure \ref{fig_two_dim_ms_scmos} shows a stitched image reconstructed from a similar dataset acquired using the Hamamatsu C12849-111U sCMOS detector at 14.4 keV. 
The stitched images were generated using the Grid/Collection Stitching plugin \cite{fiji-2} in Fiji \cite{fiji-1}. 
For both detectors, the two-dimensional multishot measurements were performed at a resolution that allowed the images to be stitched successfully. 
These results confirm that the control system operated as intended during automated multishot acquisition. 

\subsection{Computed Laminography Data Acquisition}

\begin{figure}
\centering
\includegraphics[width=\linewidth]{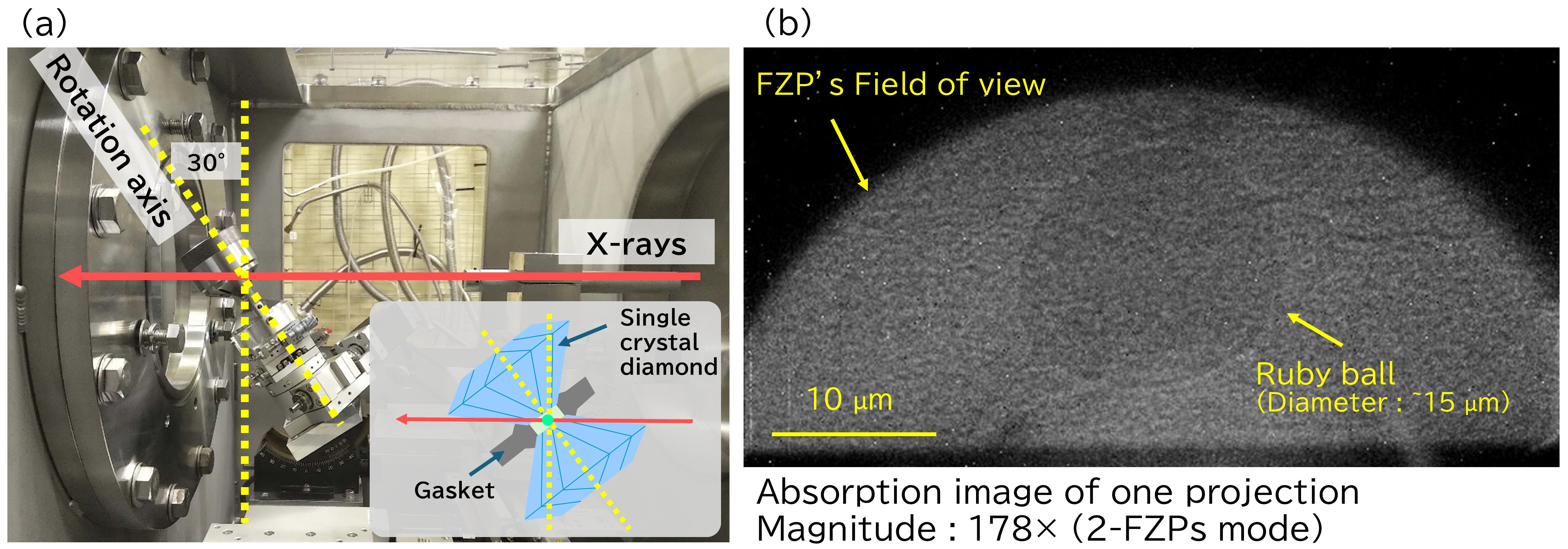}
\caption{\label{fig_lamino} Computed-laminography measurement: (a) experimental setup and (b) a single image of the small ruby ball inside the DAC, extracted from the acquired laminography dataset.}
\end{figure}

Continuous automated measurements were performed over a full rotation in \ang{0.5} increments, yielding 721 datasets, using a small ruby ball placed in a panoramic diamond anvil cell (DAC). 
Figure \ref{fig_lamino}(a) shows the computed-laminography setup. 
The measurement was performed in the 2-FZPs mode at a magnification of \(178\times\) and an X-ray energy of 9.6 keV. 
The DAC had two symmetric radial openings, each providing an unobstructed angular range of \ang{150}, for a combined angular opening of \ang{300}. 
The diamond culet diameter was 300 {\textmu}m, and the applied pressure was 10 GPa. 
The W-Re gasket was 40 {\textmu}m thick and had a hole diameter of 120 {\textmu}m. 
A ruby ball approximately 15 {\textmu}m in diameter was used as the sample, with cBN powder as the pressure-transmitting medium. The measurements were performed using the INTPIX4NA SOIPIX detector. 
Figure \ref{fig_lamino}(b) shows a single image of the ruby ball inside the DAC, extracted directly from the acquired laminography dataset. 
The complete dataset was acquired without missing data points. 
These results confirm that the proposed control system operated consistently with its intended design during automated laminography acquisition. 

\subsection{System Reliability and Operational Performance}

During the approximately one-week commissioning period, several thousand DAQ-state transitions were executed with no CCDC-related failures. 
The round-trip time for transmitting a CCDC command through the STARS framework was typically on the order of a few milliseconds, depending on the network configuration. 
These results indicate that the proposed control architecture provides sufficient responsiveness and operational reliability for routine beamline measurements.

\section{Conclusions}

A unified control system based on the STARS framework and CCDC, a common data acquisition (DAQ) state model and command set for detector control, was developed as a model implementation for the 2-FZPs zooming optics installed at the PF AR-NE1A beamline. 
This control system demonstrates modular extensibility through the STARS framework and detector interoperability through CCDC. 
It was also designed as a practical system that facilitates operation of the zooming optics by beamline users. 
Demonstration measurements performed using the optical setup at the PF AR-NE1A beamline confirmed that the system operated as intended. 
The system will be made available to users in the near future, and further improvements will be implemented based on user feedback. 
The STARS-based integrated optical-control system and CCDC developed in this study can be readily applied to other systems. 
By consolidating detector operations into a common set of DAQ transition commands, CCDC reduces the need for detector-specific modifications to measurement procedures and facilitates the integration of additional detector systems. 
Future work will include developing a generic framework for CCDC-compatible detector-control clients using detector-specific command-definition files (e.g., JSON or XML formats) and automatic client generation based on device identifiers. 
These developments are expected to further improve detector integration and plug-and-play usability. 
The STARS-based integrated optical-control system and CCDC may also support future multiprobe measurements involving X-rays, muons, neutrons, and lasers. Their application is also anticipated in planned next-generation facilities, such as the Multi Quantum-Beam Facility with Superconducting Linac for Material-Quantum-Life Sciences \cite{mblinq}. 

\section{Acknowledgements}

This study was conducted at the AR-NE1A beamline of the Photon Factory at KEK under Proposal Nos. 2021PF-S001, 2024PF-G006, 2025PF-G004, and 2025PF-G016. 
It was supported by JSPS KAKENHI Grant Nos. 21K14174, 24K00740, and 25K22028. 

\printcredits

\bibliographystyle{model1-num-names}

\bibliography{zooming-refs.bib}



\end{document}